\documentclass[12pt]{article}

\usepackage{hyperref}
\usepackage{tgtermes}
\usepackage{graphicx}
\usepackage{capt-of}
\usepackage[export]{adjustbox}

\newcommand{\beginsupplement}{%
        \setcounter{table}{0}
        \renewcommand{\thetable}{S\arabic{table}}%
        \setcounter{figure}{0}
        \renewcommand{\thefigure}{S\arabic{figure}}%
}

\usepackage[right]{lineno}

\renewcommand{\beginsupplement}{%
        \setcounter{table}{0}
        \renewcommand{\thetable}{S\arabic{table}}%
        \setcounter{figure}{0}
        \renewcommand{\thefigure}{S\arabic{figure}}%
}

\usepackage{fancyhdr}
\usepackage{tabularx}

\begin{document}

\vspace{3cm}

{
\noindent
\Large
\bf
Mapping the bacterial ways of life
}
\vspace{0.25cm}

{
\noindent
\large
Ashkaan K. Fahimipour$^{1,2,\ast}$, Thilo Gross$^{1,3,4,5}$
}

{
\footnotesize
\noindent
$^{1}$Department of Computer Science, University of California, Davis, CA, USA
}

{
\footnotesize
\noindent
$^{2}$Southwest Fisheries Science Center, Ntl. Oceanic and Atmospheric Administration, Santa Cruz, CA, USA
}

{
\footnotesize
\noindent
$^{3}$Helmholtz Instutitut for Functional Marine Biodiversity, Oldenburg, DE
}

{
\footnotesize
\noindent
$^{4}$Alfred Wegener Institute, Helmholtz Centre for Marine and Polar Research, Bremerhaven, DE
}

{
\footnotesize
\noindent
$^{5}$University of Oldenburg, Oldenburg, DE
}

{
\footnotesize
\noindent
$^\ast$To whom correspondence should be addressed;
e-mail:
ashkaan.fahimipour@noaa.gov
}

\date{}

\vspace{0.5cm}

{
\noindent
\bf
The rise in the availability of bacterial genomes defines a need for synthesis: abstracting from individual taxa, to see larger patterns of bacterial lifestyles across systems. A key concept for such synthesis in ecology is the niche, the set of capabilities that enables a population's persistence and defines its impact on the environment. The set of possible niches forms the niche space, a conceptual space delineating ways in which persistence in a system is possible. Here we use manifold learning to map the space of metabolic networks representing thousands of bacterial genera. The results reveal a metabolic niche space with a complex branching geometry, whose branches constitute major strategies spanning life in different habitats and hosts.
We further demonstrate that communities from similar ecosystem types map to characteristic regions of this new functional coordinate system, permitting ecological descriptions of microbiomes in terms of large scale metabolic roles that may be filled.
}

\section*{Main Text}

\subsection*{Introduction}
It has been pointed out that a key to understanding the rules of life in ecological communities is to understand the structure of the niche space, the sets of ecological strategies that enable populations to grow and reproduce in an ecosystem \cite{hutchinson1957cold, macarthur1972coexistence, chase2003ecological, holt2009bringing, winemiller2015functional, pianka2017toward}.
Conceptual theories envision the niche space as an $n$-dimensional geometrical shape \cite{hutchinson1957cold, blonder2014n} where each dimension is spanned by variables representing, often nonlinear combinations of salient traits or environmental features \cite{hoogenboom2009defining,porter2009size,kraft2015plant,blonder2018hypervolume}.
Empirical characterizations of the niche space have so far been conducted with a focus on individual groups of macrobiotic species, where different data analysis methods have been used to organize sets of functional traits that associate with major ecological roles in a system \cite{blonder2018hypervolume}; included are lizards \cite{winemiller2015functional},
beetles \cite{stevenson1982hutchinsonian, inward2011local},
neotropical fish \cite{pianka2017toward},
and terrestrial vascular plants \cite{kraft2015plant, diaz2016global}.

Bacteria are an attractive target for examining niche-based theories in ecology \cite{green2008microbial, fierer2007toward, horner2006phylogenetic, lennon2012mapping, fisher2017variable} as many of the relevant traits, such as the ability to metabolize certain substrates or synthesize molecules that mediate ecological interactions, are biochemical in nature \cite{prosser2007role, borenstein2008large}. Hence they can be inferred from genomes, providing plentiful data to map the niche space on a grander scale. To operationalize the bacterial niche space we say that the sets of biochemical reactions encoded by genomes represent feasible \textsl{metabolic strategies} of extant microorganisms \cite{winemiller2015functional,humphries2014metabolic, chase2011ecological}.
Together the strategies span a metabolic niche space \cite{hutchinson1957cold}: the space of metabolic capabilities that populations may deploy to survive.

Ecological niches are thought to comprise complex nonlinear functions of multiple measurable traits \cite{winemiller2015functional, kraft2015plant, blonder2018hypervolume, d2016challenges}. A central challenge in modeling the niche is thus to identify composites of traits that map to interpretable ecological roles, or the `soft properties' \cite{barter2019manifold} that summarize an organism's functional capabilities.
A powerful analysis method for meeting this challenge is offered by diffusion maps \cite{coifman2005geometric, coifman2006diffusion}. 
This mathematically simple manifold learning method leverages the physical relationship between diffusion processes and geometry to define a new coordinate system for a dataset, where the axes, or variables, are nonlinear composites of its major features. While the mathematical procedure does not provide an interpretation of these variables, our analyses show that they correspond to meaningful metabolic strategies. This offers a potential bridge between ecological niche theories and information that is readily accessible from annotated bacterial genomes.

Here we use the diffusion map to construct a trait-based functional coordinate system spanning the bacterial metabolic niche space. As a compact prediction of metabolism, we generated genome-scale metabolic networks \cite{borenstein2008large, machado2018fast} for a representative species from all unique bacterial genera in the NCBI RefSeq \cite{pruitt2006ncbi} release 92 database ($N = 2,621$ genera). 
Each representative network was mapped to a point in a 7,769-dimensional discrete space, where axes indicate the presence or absence of predicted metabolic `traits' given by unique chemical substrate-product pairs (i.e.~directed edges in the collection of metabolic networks). Although a complete picture of bacterial metabolism from genomic data is not yet possible, this array captures the major biochemical capabilities \cite{mendes2016mminte} for a large fraction of known bacterial genera, and served as input to the diffusion map algorithm.

\subsection*{Results}
The diffusion map finds new variables that reflect nonlinear combinations of metabolic capabilities and returns them in the order of their importance \cite{coifman2005geometric, coifman2006diffusion}. Each variable assigns coordinate entries to the genomes that can then be used to order genera, from the most negative to the most positive entries, along different niche dimensions. Dimensions can then be interpreted by analyzing the strategies of taxa near the extrema of the orderings \cite{barter2019manifold}, corresponding to large positive or negative (i.e. far from zero) variable entries.

\begin{figure}[t!]
\centering
\includegraphics[width=0.7\linewidth]{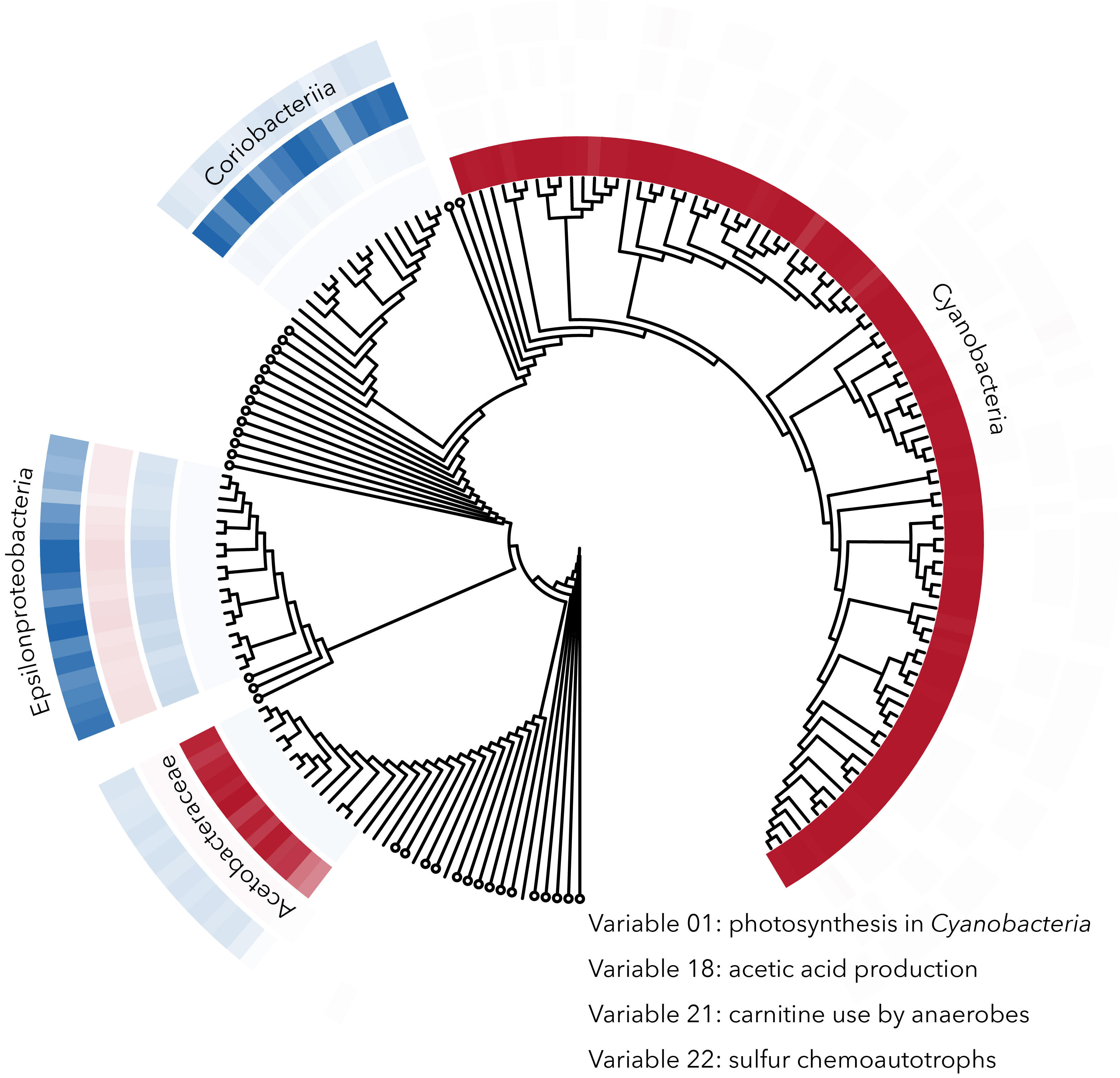}
\caption{
\footnotesize
\linespread{1}\selectfont{
The diffusion map identifies variables describing discrete yes-or-no bacterial metabolic strategies. Variable entries for each genome are visualized as colored tiles near the tips of a phylogenetic tree. Large negative or positive values (saturated reds and blues) indicate strong overlap with the focal strategy, whereas white indicates an absence of these capabilities. Circles are collapsed clades with near-zero entries in each of the four example variables. Clades receiving large negative or positive entries in any of the four example variables are expanded and annotated. The near-absence of semi-saturated tones indicates that deploying the strategies represented by these variables is approximately a yes-or-no decision.
}
}
\label{fig:fig_1}
\end{figure}

\subsubsection*{Sharp differences delineate some metabolic strategies}
The most important variable identified by the diffusion map, variable 1, separates the metabolic strategies of photosynthetic \textit{Cyanobacteria} from those of all other taxa: the 108 cyanobacterial genomes in the dataset are assigned low values (i.e.~negative numbers with large magnitudes) in variable 1, while all others have values that are close to zero (Fig.~1; Fig.~S1A). To confirm that this variable detects cyanobacterial photosynthetic activity, we identified metabolites that were over-represented in the metabolic networks of genera receiving far-from-zero entries in variable 1 (see \textsl{Methods}). This revealed an enrichment of 2-Phosphoglycolate, which is involved in essential photorespiratory pathways in photosynthetic organisms \cite{eisenhut2008photorespiratory}; ribulose-1,5-bisphosphate (RuBP), used for carbon fixation from RuBisCO during photosynthesis; cyanophycin, a unique nitrogen reserve polymer \cite{watzer2018cyanophycin}; and sucrose 6-phosphate, which catalyzes the final step in sucrose biosynthesis in \textit{Cyanobacteria} \cite{fieulaine2005structure}, confirming that the variable indicates the extent to which \textit{Cyanobacteria} fix carbon through photosynthesis (Fig.~1; Table S1).

The sharp differences in variable 1 show that this photosynthetic lifestyle is a discrete yes-or-no metabolic strategy where little middle ground exists. The diffusion map defines further variables that indicate such discrete clade-level capabilities (Fig.~1), so-called `localized' variables \cite{nyberg2015mesoscopic}, including capabilities associated with acetic acid production \cite{komagata2014family} (variable 18), carnitine use for stress tolerance among anaerobic animal associates \cite{meadows2015carnitine} (variable 21), and chemolithoautotrophic or sulfur-oxidization strategies deployed by $\epsilon$-proteobacteria near anoxic marine sediments and sea vents (variable 22).

\subsubsection*{Contrasting the major strategies of host associates to life in soils and oceans}
Some variables identified by the diffusion map analysis span a continuous spectrum of strategies, which align with major taxonomic classes. The most important of these are variables 2, 3, and 4, which contrast different putative metabolic strategies encoded by relatively large proportions of the analyzed genomes (Fig.~2; Fig.~S1B). For instance, variable 2 identifies major differences in predicted strategies among host-associated \textit{$\gamma$-proteobacteria} and soilborne \textit{Actinobacteria}. Close relatives of pathogenic \textit{Enterobacter}, \textit{Franconibacter}, and \textit{Buttiauxella} species \cite{kampfer1997characterization} score the lowest (i.e.~most negative) values (Figs.~2A and 2B). Metabolic capabilities associated with these taxa include the synthesis of membrane phospholipid precursors common in \textit{$\gamma$-proteobacteria} like CDP-diacylglycerol \cite{parsons2013bacterial} and phosphatidylethanolamine, which may be involved in bacterial adhesion to host cells \cite{foster1999phosphatidylethanolamine}; and the ability to metabolize uncommon sugars like L-lyxose \cite{mayer2005hexose} (Table~S2). At the opposite end, we find primarily Gram-positive soil organisms from the \textit{Microbacteriaceae}, \textit{Beutenbergiaceae}, and \textit{Micrococcaceae} \cite{reimer2019bac} (Figs.~2A and 2B). Among the most correlated capabilities for species near this extremum are the synthesis of decaprenyl diphosphate, a key component of cell wall biosynthesis in some taxa \cite{kaur2004decaprenyl}; and compounds related to the synthesis of thiol and bimane derivatives, which can function as defenses against alkylating agents, oxygen stress, and antibiotics \cite{newton2008biosynthesis} (Table~S3).

\begin{figure*}[t!]
\centering
\includegraphics[width=0.9\textwidth]{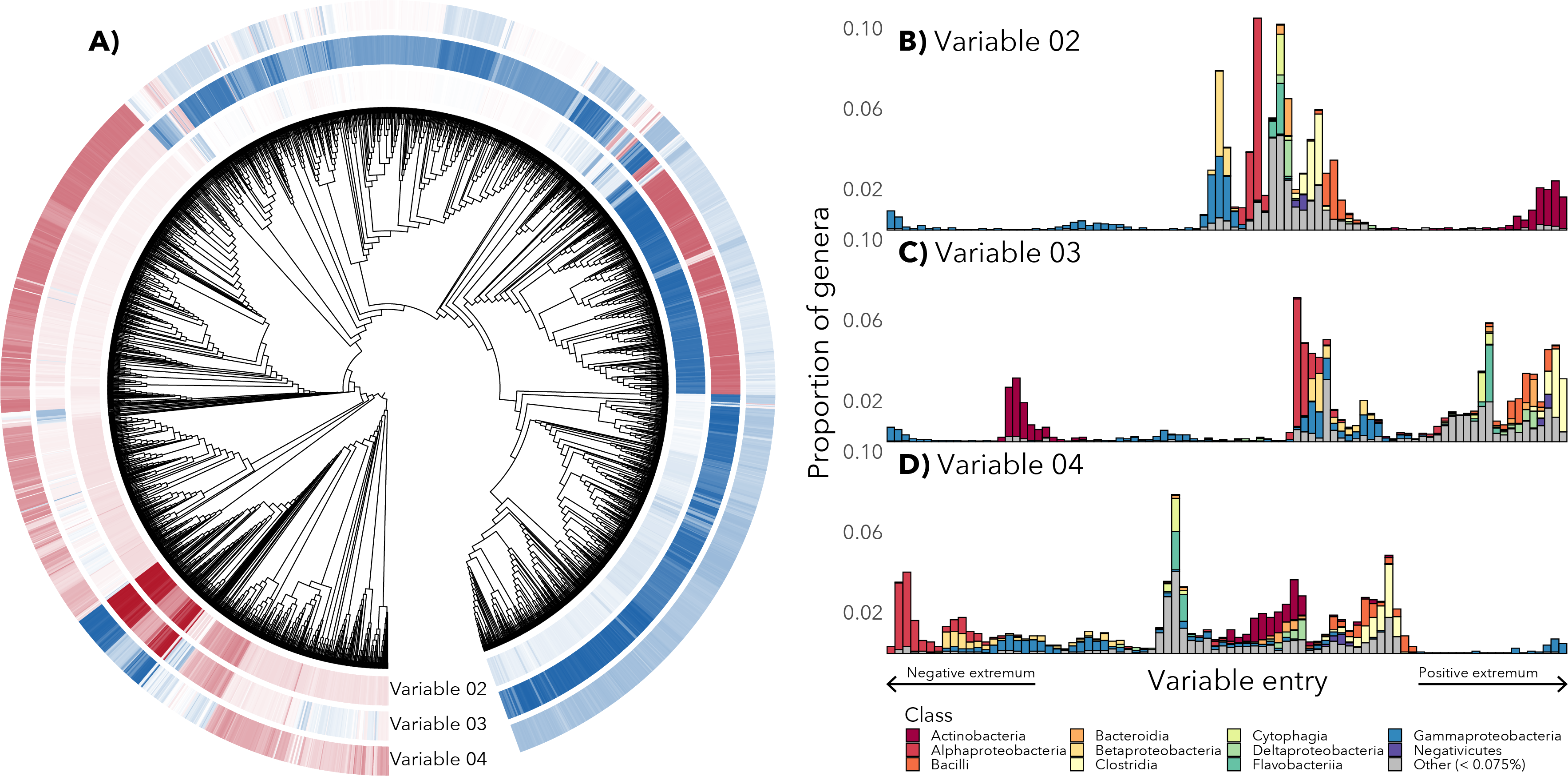}
\caption{
\footnotesize
\linespread{1}\selectfont{
A broad spectrum of class-level capabilities indicated by variables 2, 3, and 4. \textbf{A)} Variable entries for each genome are shown as tiles near the tips of a phylogenetic tree. Darker red and blue tiles mark genomes receiving larger (in magnitude) negative and positive variable entries. \textbf{B)} The ordering of taxa defined by variable 2 entries, from negative to positive (left to right). The taxonomic compositions corresponding to variable entries are shown for each of 100 equally spaced bins.
\textbf{C)} The ordering of taxa defined by variable 3 entries. \textbf{D)} The ordering of taxa defined by variable 4 entries. The wide variety of different values of these variables indicates a gradual shift in metabolic capabilities.
}
}
\label{fig:fig_2}
\end{figure*}

The \textit{$\gamma$-proteobacteria} genera that received the lowest entries in variable 2 also constituted the negative extremum of variable 3, and the positive extremum of variable 4 (compare Figs.~2a-2d), signifying a multiway contrast between these \textit{$\gamma$-proteobacteria} taxa and at least 3 other taxonomic classes. At the positive end of variable 3, we find taxa representing mammal- and bird-associated \textit{Clostridia}, \textit{Tissierellia}, \textit{Erysipelotrichia} and \textit{Bacilli} \cite{reimer2019bac}. Characteristic metabolites of these genera include components of the Wood-Ljungdahl pathway \cite{wei2017structural}, enabling the use of hydrogen as an electron donor; and indole, a signalling molecule that has been shown to modulate host inflammation and interspecific competition in human gastrointestinal tracts \cite{darkoh2019clostridium} (Table S4).
Our interpretation is that variable 3 identifies different potential strategies for successfully colonizing and weathering stress and interspecific competition in animal hosts.

The species that score the lowest (i.e.~most negative) values in variable 4 are epipelagic and marine \textit{Rhodobacterales}, \textit{Rhizobiales}, and \textit{Rhodospirillales} that are capable of utilizing a broad spectrum of carbon sources \cite{luo2015divergent}. Here the most significant metabolic reactions are all involved in the L-2-aminoadipate pathway of lysine synthesis \cite{kanehisa2000kegg} and the production of L-pipecolic acid (Table S5), pointing to a strategy for growth under high-salt conditions \cite{neshich2013genome}. Our interpretation is that this variable traces a range of strategies spanning a generalist lifestyle in oceans to associations with terrestrial hosts.

Host-microbe interactions also feature in variables 8 and 10, which highlight endosymbionts and endoparasites with the smallest genomes in our dataset. We found that the lowest values of variable 8 coincided with animal- and plant-associated \textit{Tenericutes} \cite{reimer2019bac}, as well as candidate genera like \textit{Tremblaya} and \textit{Sulcia}, that are associated with insect bacteriocytes \cite{chang2015complete, lopez2015link}. Among the top 10 markers of taxa scoring low values in variable 8 include the predicted uptake \cite{borenstein2008large} of key amino acids such as L-histidine, L-arginine, L-isoleucine, L-valine, L-lysine, and L-leucine (Table S6). The negative extremum of variable 10 features obligate endoparastites and close relatives of opportunistic pathogens, including putative animal- and arthropod-associates of the \textit{Pasteurellaceae}, \textit{Erwiniaceae}, \textit{Morganellaceae}, and \textit{Rickettsiaceae} \cite{reimer2019bac}. Similarly to variable 8, metabolic network features that distinguished this group include the predicted uptake of L-histidine, L-arginine, L-threonine, L-isoleucine, L-glutamine, and L-lysine (Table S7). Together, these variables indicate that one widespread strategy for life in close association with animal or plant cells is a reliance on host-derived essential and non-essential amino acids \cite{dale2006molecular}.

\subsubsection*{Phylogenetic relatedness is a rough indicator of ecological similarity}
The first several diffusion variables identify characteristic capabilities that discriminate between major taxonomic classes comprising many representative genera. To assess the overall relationship between metabolic similarity and phylogenetic relatedness we computed the correlation between pairwise inter-genome metabolic distances in diffusion space, and cophenetic distances on the phylogenetic tree (see ref.~\cite{nadler2006diffusion} for a detailed discussion of diffusion distances). Here a positive correlation suggests that closely-related taxa deploy similar metabolic strategies on average. 

The Pearson correlation between distance matrices was positive but exhibited a small coefficient (Fig.~3A; Mantel test, $r = 0.273$, $P < 0.001$), marking a weak association between predicted metabolic capabilities and phylogenetic relatedness. While it is not surprising that phylogenies contain information on the potential ecological roles of microorganisms \cite{langille2013predictive, louca2018function, douglas2019picrust2}, a visualization of this relationship highlights a caveat: a large range of diffusion distances are observed for most given cophenetic distances between genome pairs (Fig.~3A). This high degree of variance can be explained by the presence of diffusion variables that deviate from basic contrasts between major taxonomic groups (e.g.~Fig.~2), including some that differentiate between closely-related taxa (Fig.~S1C), and those that show similar strategies among distantly-related taxa (Fig.~3B), potentially reflecting metabolic niche convergence \cite{pianka2017toward} or horizontal gene transfer.

\begin{figure*}[t!]
\centering
\includegraphics[width=0.9\textwidth]{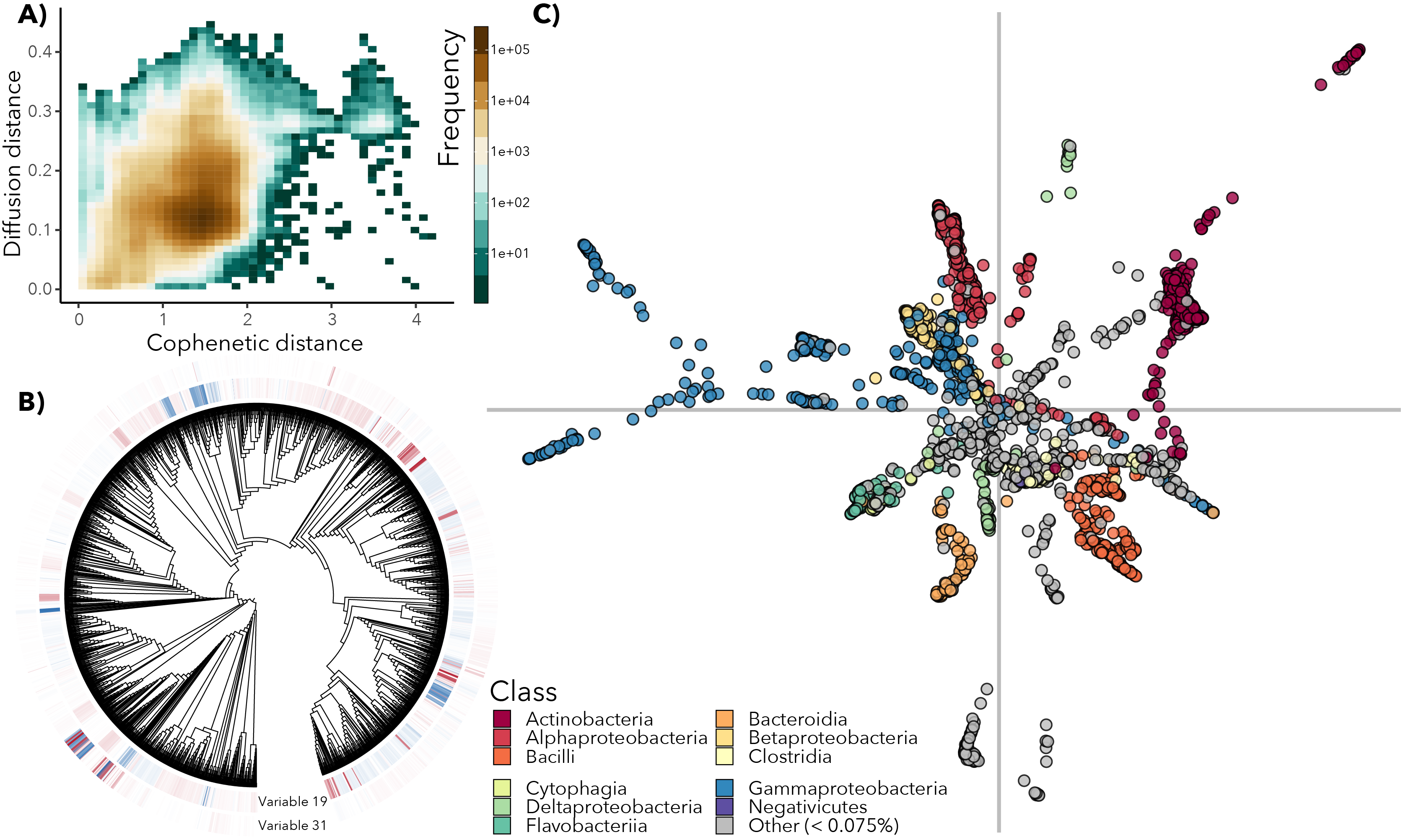}
\caption{
\footnotesize
\linespread{1}\selectfont{
Metabolic and phylogenetic similarities are roughly correlated. \textbf{A)} The relationship between distance in diffusion space, and cophenetic distance along branches of the phylogenetic tree between genome pairs. A large range of diffusion distances are observed for most given cophenetic distances. (\textbf{B}) Some variables such as 19 show similarities across the tree, with similar functional capabilities shared by remotely related taxa (similar colors in distal parts of the tree). Others such as variable 31 highlight differences in closely related taxa, corresponding to the appearance of large positive and negative values (dark blue and red shades) in close proximity on the tree. \textbf{C)} A 2-dimensional embedding of all diffusion variables \cite{moon2017phate}, where individual genomes (points) are colored by taxonomic class. Axes mark (0, 0) in the arbitrary coordinate system. This confirms the coarse alignment between phylogeny and the niche.
}
}
\label{fig:fig_3}
\end{figure*}

These examples demonstrate that diffusion variables provide dozens or possibly hundreds of meaningful coordinates that trace the space of bacterial metabolic strategies. Using a procedure proposed by Moon et al. \cite{moon2017phate} we combined diffusion variables in a low-dimensional visualization of the strategy space (Fig.~3C; Fig.~S1). This embedding recapitulates the result that phylogenetic relatedness offers only a coarse marker of predicted ecological similarity, corresponding to the appearance of representatives from multiple taxonomic classes in close proximity to one another in the metabolic niche space.

The 2-dimensional embedding of diffusion variables also suggests that the niche space exhibits a filamentous geometry with multiple quasi one-dimensional branches rising out of a common core. This geometry conceptually contrasts to both Hutchinson's original idea of the niche space as a solid hypervolume \cite{hutchinson1957cold}, and modern ideas which postulate that sets of functional traits may form discrete ecological clusters \cite{winemiller2015functional, pianka2017toward}. We conjecture that this filamentous structure has implications for the evolution of bacteria. For instance, the branching geometry naturally leads to a large amount of unoccupied niche space (Fig.~3C). This empty space could correspond to bacteria that have yet to be sampled, isolated, or sequenced, but it could also arise due to `forbidden' metabolisms, i.e.~combinations of capabilities that may be suboptimal or even pointless for life in Earth's ecosystems.

\subsubsection*{Microbiomes map to characteristic regions of the metabolic niche space}
Understanding the mapping from genomes to larger scale ecological strategies may prove useful for a variety of analyses \cite{green2008microbial,fierer2007toward,horner2006phylogenetic}, such as quantifying the roles of organisms or designing substrates for culturing. Perhaps more importantly it provides an ecological frame of reference for characterizing bacterial communities. For a small scale demonstration of this point we created a simple mapping between a subset of community censuses from the Earth Microbiome Project (EMP) \cite{thompson2017communal} and our diffusion space. 

\begin{figure*}[t!]
\centering
\includegraphics[width=1\textwidth]{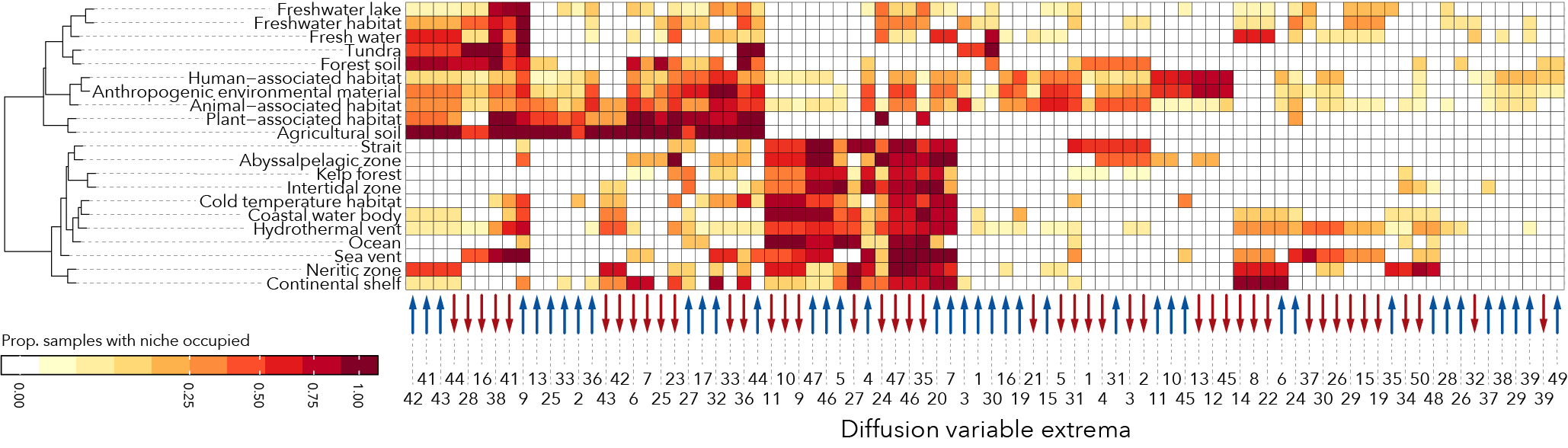}
\caption{
\footnotesize
\linespread{1}\selectfont{
Bacterial communities across different ecosystem types (rows) in the Earth Microbiome Project \cite{thompson2017communal} exhibit characteristic metabolic niche profiles. Columns correspond to different diffusion variable extrema. Darker tiles indicate that a larger fraction of community censuses contained taxa that mapped to those extrema. Blue and red arrows along the horizontal axis denote positive and negative variable extrema respectively. A hierarchical cluster analysis groups ecosystems with similar niche profiles.
}
}
\end{figure*}

First, for each selected bacterial community census in the EMP we matched all taxa
(16S rRNA gene amplicon sequence variants) to the most closely related genome considered by our
diffusion map analysis, and retained matches that exhibited at least a 97\% sequence similarity (see \textsl{Methods}). We then determined whether EMP communities contained at least one taxon
that mapped to any of the 10 extremal genomes along any of the
first 50 diffusion variables. As a result, each microbiome sample was characterized by the presence or absence of each of the first 100 extremal metabolic strategies. These presence-absence data represent ecological characterizations for individual EMP communities. To coarse-grain further we computed the
proportions of communities from different ecosystem types that displayed the different extremal strategies. The result is a bacterial metabolic fingerprint for a type of ecosystem (Fig.~4). These fingerprints can then be used to study systematic differences in the predicted capabilities of typical community members across different ecosystem types. For example, a simple hierarchical clustering analysis of metabolic fingerprints groups different ecosystem types meaningfully together based on the strategies deployed by their constituents (Fig.~4). Visible are clear strategy sets that differentiate life in freshwater, soil, marine, and host associated systems.

\subsection*{Discussion}
Here we showed that the shape of a trait space can be systematized through manifold learning \cite{coifman2005geometric}.
The diffusion map of bacterial capabilities reveals a wealth of ecologically salient variables that span a functional coordinate system. Some show evidence of discrete capabilities such as photosynthesis in \textit{Cyanobacteria} (Fig.~1). Other strategies span a continuous space representing degrees of specialization or reliance on hosts (Fig.~2). Yet others highlight strategies for energy production or stress response, some of which differentiate closely related species (Fig.~S1C) or emerged, potentially through convergent evolution or gene transfer, in different branches of the tree of life (Fig.~3B). 

The diffusion variables provide a physical method for organizing the genomic information that continues to emerge, in a way that reveals both larger scale geometries and finer details compared to alternative embedding methods \cite{coifman2005geometric, moon2017phate} (\textsl{Supplementary Information}; Figs.~S1-S4). From the perspective of microbial systems, diffusion distances in trait space (e.g., Fig.~3A) provide a powerful proxy for ecological similarities that can complement insights from current phylogenetic methods \cite{douglas2019picrust2, lozupone2005unifrac}. Traits used to calculate diffusion distances need not be derived from metabolic reconstructions of whole genomes as in the present analysis, but could comprise functional information identified, for instance, through species-level profiling \cite{franzosa2018species} of metagenomic or metatranscriptomic shotgun sequencing data. From an ecological point of view the present analysis constitutes the most extensive mapping of a niche space geometry so far, and facilitates the application of ecological theories to data describing bacterial communities. 

Our analysis focused largely on the bacteria's capabilities to catalyze steps in primary metabolism. Even within the realm of primary metabolism the genes reveal only the set of theoretical capabilities encoded by genomes, conceptually analogous to the fundamental niche concept \cite{hutchinson1957cold} in ecology. Hence our analysis ignores uncharacterized parts of secondary metabolism, behavior, regulation, and trophic interactions \cite{shiratori2019phagocytosis}. For any other group of organisms such a limited analysis would be mostly meaningless, however due to the diversity of metabolic capabilities in bacteria it reveals a rich and complex functional coordinate system (Fig.~3C). As our understanding of genomic data advances, deeper insights into secondary metabolism are bound to become available, providing an even more detailed picture of the metabolic niche space. Moreover, we envision that with future transcriptomic data, manifold learning methods could also map the realized niche \cite{hutchinson1957cold} (the metabolic strategies that are deployed under a given set of conditions) bringing our understanding of ecology in complex microbial communities closer to the biochemical level.

\clearpage
\subsection*{Methods}
\subsubsection*{Metabolic networks}
Genomes were obtained from the National Center for Biotechnology Information (NCBI) RefSeq \cite{pruitt2006ncbi} database (accessed on 2019 March 20). We first obtained the `representative,' `reference,' `complete,' `contig,' and `scaffold' sets and reduced these to a set of genus-level representatives using the following sampling procedure. We first selected a random representative genome for all unique genera in the combined `representative' and `reference' sets. Novel genera in the remaining RefSeq categories, that were not already represented in the `reference' and `representative' sets, were then appended to the set in the same way, for a total $N = 2,621$ genomes. Metabolic models were constructed for the selected genome assemblies using the \texttt{CarveMe} reconstruction algorithm \cite{machado2018fast}, that starts with a universal bacterial metabolic model comprising known biochemical reactions in the BiGG Models \cite{king2015bigg} database and generates genome-specific reaction sets by paring those without genomic support. Finally, metabolic models' cytoplasmic compartments were summarized as metabolic networks --- directed graphs in which nodes are chemical metabolite compounds and directed edges link substrates to products \cite{borenstein2008large}.

\subsubsection*{Phylogenetic tree generation}
Phylogenetic trees were used to visualize metabolic differences between taxa, and were constructed using the \texttt{GToTree} pipeline \cite{lee2019gtotree} with the ``universal'' protein set defined by Hug et al. \cite{hug2016new}.
\texttt{GToTree} identifies target genes with \texttt{HMMER3} \cite{eddy2011accelerated}, aligns them with \texttt{MUSCLE} \cite{edgar2004muscle}, and trims alignments with \texttt{trimAl} \cite{capella2009trimal}. Trees were generated from the aligned and concatenated gene sets using \texttt{FastTree} \cite{price2010fasttree}, and visualized using iToL \cite{letunic2019interactive}.

\subsubsection*{Identifying associated metabolites}
We sought to identify metabolites that were overrepresented in the metabolic networks of taxa, that were themselves assigned extreme entries along diffusion map variables. This was accomplished using a permutational analysis analogous to the gene set enrichment analysis, GSEA \cite{subramanian2005gene}. Genome rankings were provided by the orderings specified by each diffusion variable. Enrichment analyses were accomplished for the preranked sets using the \texttt{fgsea} library in R, with an FDR-adjusted $P$-value $< 0.05$ used as the threshold for retaining metabolites associated with taxa that map to variable extrema.

\subsubsection*{Diffusion map procedure}
Diffusion mapping \cite{coifman2005geometric,coifman2006diffusion} was performed using the algorithm described by Barter \& Gross \cite{barter2019manifold}. Briefly, the method involves (i) calculating an affinity matrix describing euclidean similarities among the $k$-nearest neighbors for samples in a dataset, (ii) interpreting this as a weighted adjacency matrix, and (iii) computing the corresponding row-normalized Laplacian matrix. The eigenvectors of the Laplacian represent new diffusion variables describing important variation in the dataset. This method is nearly parameter-free, with only a single choice for the value of $k$. Here, we consider $k = 10$, although the results presented above were insensitive to the choice of $k$. The first (i.e.~most important) variable is given by the eigenvector corresponding to the smallest non-zero eigenvalue, then the second smallest eigenvalue, and so on. We provide an R \cite{rcoreteam} implementation of this procedure at \href{https://github.com/AshkaanF/diffusion\_maps}{https://github.com/AshkaanF/diffusion\_maps}.

\subsubsection*{Mapping environmental samples to diffusion space.}
We obtained the preprocessed `emp\_deblur\_150bp.subset\_2k.rare\_5000' dataset, describing a subset of the environmental 16S rRNA gene sequences from the Earth Microbiome Project \cite{thompson2017communal}, EMP, accessed via \href{ftp://ftp.microbio.me/emp/}{ftp://ftp.microbio.me/emp/}. Communities from the EMP were mapped to our diffusion space using the following procedure: First, we generated a \texttt{BLAST} \cite{altschul1990basic} reference database of predicted 16S rRNA gene sequences for our set of RefSeq genomes using \texttt{barrnap} (\href{https://github.com/tseemann/barrnap}{https://github.com/tseemann/barrnap}) to identify and retain the first instance of this ribosomal gene. The \texttt{DECIPHER} library in R was used to align sequences \cite{wright2016using}. We then conducted a \texttt{BLAST} sequence similarity search \cite{altschul1990basic} to match denoised sequence variants present in each EMP sample to the \texttt{BLAST} database and retained the top hits. Niches -- operationally defined as the strategies describing the 10 taxa with the highest (positive) and lowest (negative) entries along each diffusion variable -- were said to be `occupied' by taxa in an EMP community census if at least one detected sequence variant exhibited a 97\% or greater rRNA gene sequence similarity to any of the extremal genomes. The results of this procedure were summarized as plots of the proportion of samples within each EMP `env\_feature' category satisfying this criterion. Groupings of similar ecosystem types were accomplished using the Ward linkage method \cite{ward1963hierarchical} for hierarchical clustering.

\bibliographystyle{naturemag}
{\footnotesize
\bibliography{ncommbib}
}

\subsubsection*{Acknowledgments}
We thank Jonathan~A.~Eisen and James~P.~O'Dwyer, for comments. A.K.F and T.G. were supported by the Dept.~of Computer Science at the University of California, Davis. A.K.F is supported by a National Research Council associateship.

\subsubsection*{Authors contributions} 
A.K.F and T.G.~conceptualized the study, wrote the manuscript, and contributed analyses.  A.K.F.~contributed computer code.

\subsubsection*{Competing interests}
The authors declare no competing interests.

\subsubsection*{Data and materials availability} 
R scripts and bacterial genome accession numbers have been made available at \href{https://github.com/AshkaanF/diffusion_maps}{https://github.com/AshkaanF/diffusion\_maps}.

\clearpage
\section*{Supplementary Information}
\beginsupplement

{
\large
\noindent
Mapping the bacterial ways of life
}

{
\noindent
\small
Ashkaan K. Fahimipour$^{1,2,\ast}$, Thilo Gross$^{1,3,4}$
}

{
\footnotesize
\noindent
$^{1}$Department of Computer Science, University of California, Davis, CA, USA
}

{
\footnotesize
\noindent
$^{2}$Southwest Fisheries Science Center, Ntl. Oceanic and Atmospheric Administration, Santa Cruz, CA, USA
}

{
\footnotesize
\noindent
$^{3}$Helmholtz Instutitut for Functional Marine Biodiversity, Oldenburg, DE
}

{
\footnotesize
\noindent
$^{4}$Alfred Wegener Institut for Marine and Polar Research, Bremerhaven, DE
}

{
\footnotesize
\noindent
$^\ast$To whom correspondence should be addressed;
e-mail:
ashkaan.fahimipour@noaa.gov
}
\vspace{0.5cm}

\noindent
Commonly used dimensionality reduction methods in biology include principal component analysis (PCA), multidimensional scaling (MDS), and t-distributed stochastic neighbor embedding (t-SNE) \cite{maaten2008visualizing}. These methods provide valuable insights across a variety of applications (see ref. \cite{moon2017phate} for a comparison of embedding methods with high-throughput sequencing data). Here we present examples of their limitations in the context of metabolic trait space reconstruction. Consistent with prior results \cite{moon2017phate}, we observe that PCA identifies some important global contrasts but misses fine details, that MDS is highly sensitive to outliers, and that neighbor embedding methods like t-SNE find some localized \cite{nyberg2015mesoscopic} clusters by preserving local data structures but cannot reliably identify global structures.

\subsubsection*{Desirable properties of an ecological coordinate system}
An effective ecological coordinate system for high-dimensional trait data would ideally preserve important local and global features of the dataset.
Here, we demonstrate by way of example that diffusion-based methods learn both local and global structures, while popular methods are deficient in one or both of these desirable properties \cite{moon2017phate}.

\begin{figure*}[t!]
\centering
\includegraphics[width=0.9\textwidth]{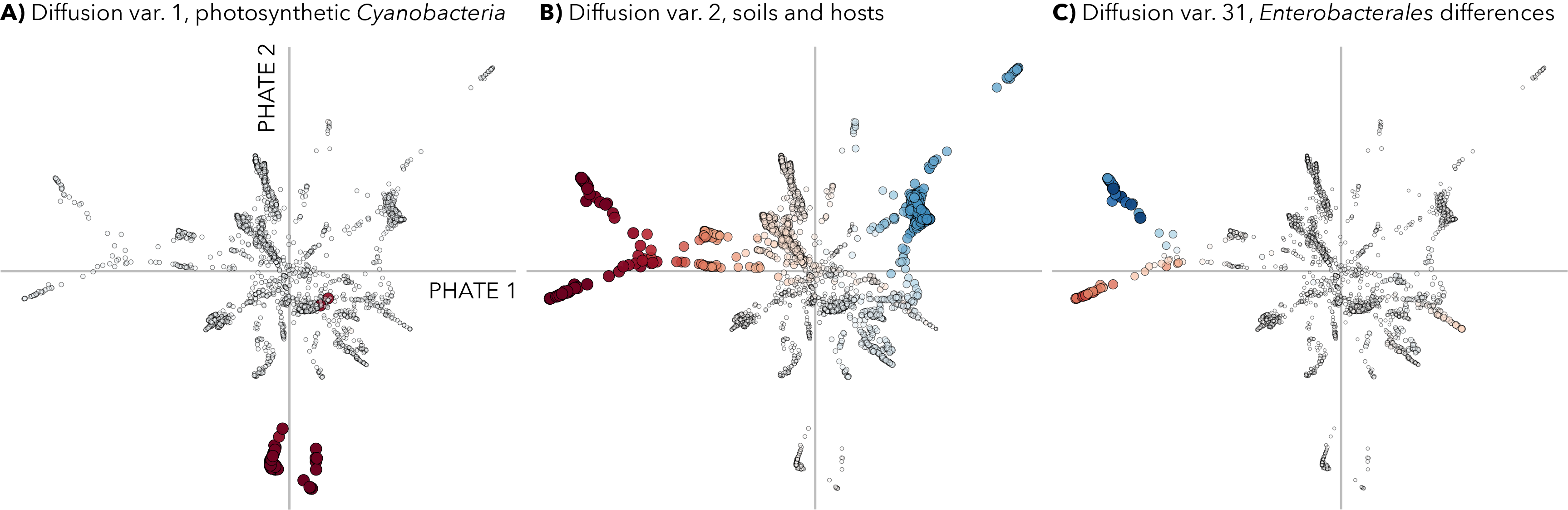}
\caption{
\footnotesize
\linespread{1}\selectfont{
Two-dimensional embedding of diffusion variables, computed using the `PHATE' algorithm \cite{moon2017phate}. Points are individual genomes colored by their assigned entries in a specific diffusion variable. Dark shades of red and blue correspond to small (i.e., the most negative) and large (positive) variable entries; white points are near zero. Axes mark (0, 0) in the coordinate system. \textbf{A)} Diffusion variable 1, identifying photosynthetic capabilities in \textit{Cyanobacteria} (dark red points). \textbf{B)} Diffusion variable 2, which identifies differences between soil-associated \textit{Actinobacteria} (dark blue) and host-associated \textit{$\gamma$-proteobacteria} (dark red). \textbf{C)} Diffusion variable 31, which finds a functional split among close relatives in the \textit{Enterobacterales}.
}
}
\label{fig:fig_s1}
\end{figure*}

\begin{figure*}[b!]
\centering
\includegraphics[width=0.9\textwidth]{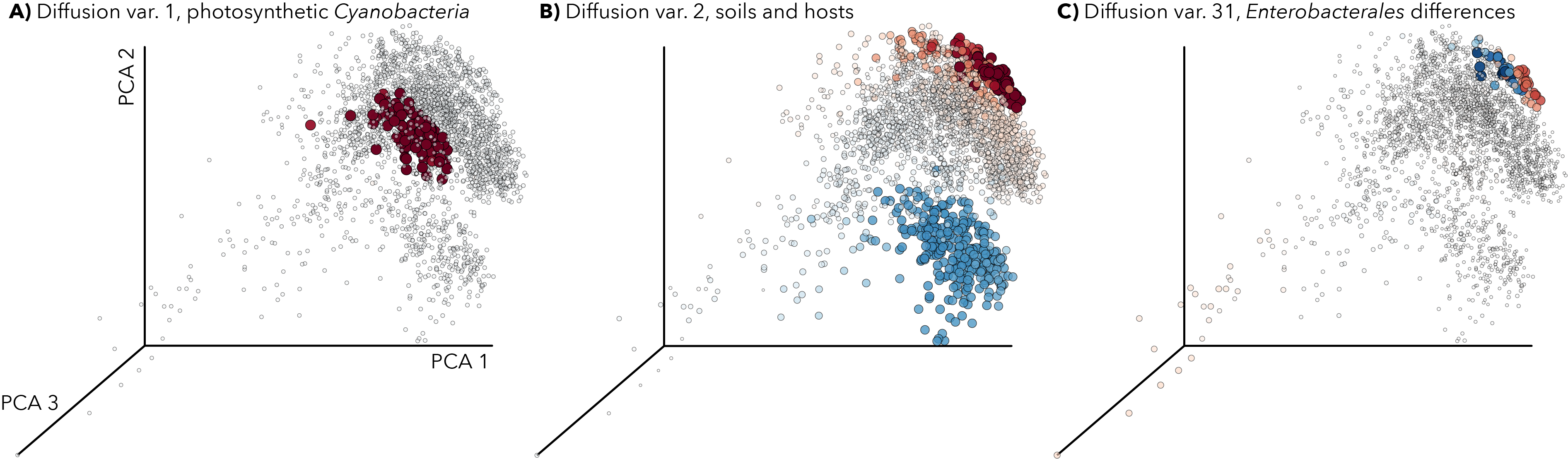}
\caption{
\footnotesize
\linespread{1}\selectfont{
Three-dimensional embedding of data points using principal component analysis. Points are individual genomes colored by their assigned entries in a specific diffusion variable (see Fig.~S1 for a description). \textbf{A)} Diffusion variable 1, \textbf{B)} diffusion variable 2, and \textbf{C)} diffusion variable 31.
}
}
\label{fig:fig_s2}
\end{figure*}

We consider three examples of types of functional `soft properties' \cite{barter2019manifold} identified by the diffusion map and supported by enrichment analysis (Tables~S1-S7) \cite{subramanian2005gene}: an example of capabilities that uniquely distinguish a group from all others (variable 1; carbon fixation by photosynthetic \textit{Cyanobacteria}), an example of conserved differences between major taxonomic classes (variable 2; soilborne \textit{Actinobacteria} vs. host-associated \textit{$\gamma$-proteobacteria}), and an example of major differences among close relatives (variable 31; differences among species in the \textit{Enterobacterales}).

Figure S1 shows the two-dimensional embedding of diffusion variables using the `potential of heat-diffusion for affinity-based transition embedding' (PHATE) method (also see Fig.~3C in the main text) of Moon et al. \cite{moon2017phate}. Coloring the points (genomes) by diffusion variable entries shows how the method captures both intricate (Figs. S1A, S1C) and global scale (Fig. S1B) structures in the metabolic trait data in only two dimensions.

In contrast, other methods discard important information encoded in higher dimensions. This is well-understood for linear methods like PCA that focus on explaining global variances, as this comes at the expense of distorting fine grained structures in data \cite{moon2017phate} (Fig.~S2A-S2C). In practice, this means that PCA is able to find some global contrasts in our data, like the major differences in metabolic capabilities between \textit{Actinobacteria} and \textit{$\gamma$-proteobacteria} (Fig.~S2B).  However, the method is inherently unable to capture structures that lie near nonlinear submanifolds. As a consequence, fine details like intra-class differences in metabolic traits are obfuscated (e.g., Fig~S2C).

\begin{figure*}[t!]
\centering
\includegraphics[width=0.9\textwidth]{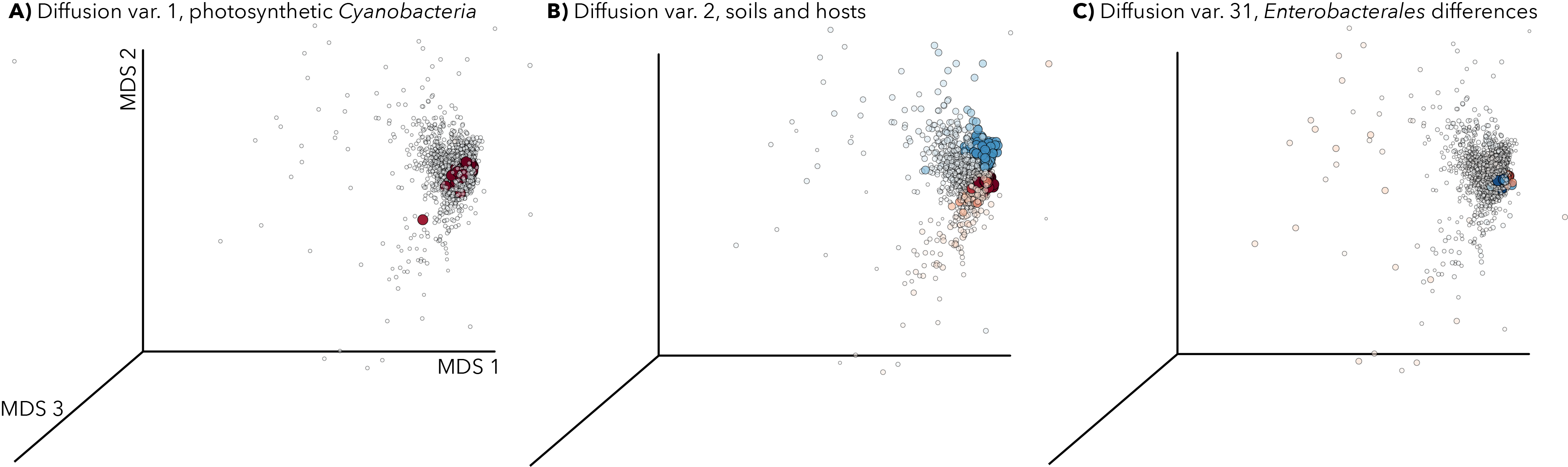}
\caption{
\footnotesize
\linespread{1}\selectfont{
Three-dimensional embedding of data points using multidimensional scaling. Points are individual genomes colored by their assigned entries in a specific diffusion variable (see Fig.~S1 for a description). \textbf{A)} Diffusion variable 1, \textbf{B)} diffusion variable 2, and \textbf{C)} diffusion variable 31.
}
}
\label{fig:fig_s3}
\end{figure*}

\begin{figure*}[b!]
\centering
\includegraphics[width=0.9\textwidth]{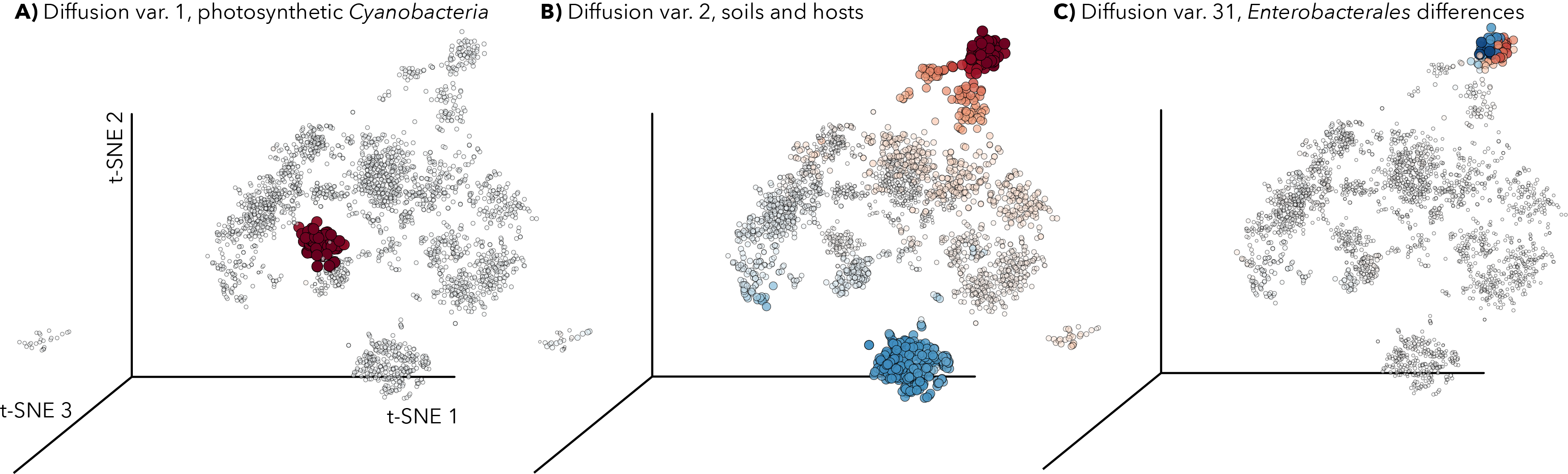}
\caption{
\footnotesize
\linespread{1}\selectfont{
Three-dimensional embedding of data points using t-distributed stochastic neighbor embedding. Points are individual genomes colored by their assigned entries in a specific diffusion variable (see Fig.~S1 for a description). \textbf{A)} Diffusion variable 1, \textbf{B)} diffusion variable 2, and \textbf{C)} diffusion variable 31.
}
}
\label{fig:fig_s4}
\end{figure*}

Another notable concern is the sensitivity of dimensionality reduction methods to noise \cite{moon2017phate}. This problem is particularly apparent in the application of multidimensional scaling to our metabolic trait dataset. Visible in the embedding is a strong sensitivity to outliers (Figs. S3A-S3C), leading to a tight cluster of genomes with very little visible structure. Within this cluster, important features are generally difficult to resolve.

The t-SNE \cite{maaten2008visualizing} method is capable of identifying localized \cite{nyberg2015mesoscopic} clusters in the data (e.g., differences in the \textit{Cyanobacteria}; Fig.~S4A) but cannot reliably capture global features of the dataset. This is because t-SNE minimizes an objective function that effectively ignores large dissimilarities between points, causing distances in t-SNE space to be mostly meaningless \cite{moon2017phate}. This leads to issues particularly in the analysis of data that are well-described by continuous or branching trajectories (e.g., cell differentiation data), as t-SNE shatters these trajectories leading to the false impression of data clusters \cite{moon2017phate}.

Even using a limited number of examples, several of the limitations of common methods for trait space reconstruction become apparent. Included are tradeoffs between the preservation of local and global data structures, sensitivity to noise, and prior assumptions about the linearity or geometry of the underlying data.

\clearpage
\subsection*{Supplementary Tables}
\begin{table}[!htbp] 
\centering
\footnotesize
\begin{tabular}{@{\extracolsep{0pt}} lccc} 
\\[-1.8ex]\hline
Metabolite & Synthesized & Enrich. Score & Adj. P\\ 
\hline \\[-1ex] 
2-Phosphoglycolate & Yes & $-1.073$ & $0.001$ \\ 
D-Ribulose 1,5-bisphosphate & Yes & $-1.071$ & $0.001$ \\ 
Cyanophycin & Yes & $-1.070$ & $0.001$ \\ 
1,5-Diaminopentane & Yes & $-1.067$ & $0.001$ \\ 
Sucrose 6-phosphate & Yes & $-1.067$ & $0.001$ \\ 
\hline \\[-1.8ex] 
\end{tabular} 
\caption{
\footnotesize
Top 5 over-represented metabolites in the metabolic networks of taxa that receive the most negative entries on variable 1. The \textsl{Enrich. Score} and \textsl{Adj. $P$} columns show the normalized `Enrichment score' and FDR-adjusted $P$-value from the enrichment analysis \cite{subramanian2005gene, sergushichev2016algorithm}. The \textsl{Synthesized} column indicates whether the network is predicted to produce the metabolite, based on its in-degree \cite{borenstein2008large}.
}
\end{table} 

\begin{table}[!htbp] 
\centering 
\footnotesize
\begin{tabular}{@{\extracolsep{5pt}} lccc} 
\\[-1.8ex]\hline 
Metabolite & Synthesized & Enrich. Score & Adj. P\\ 
\hline \\[-1ex]
Acyl-glycerophosphoethanolamine  & Yes & $-2.233$ & $0.001$ \\ 
Phosphatidylethanolamine & Yes & $-2.233$ & $0.001$ \\ 
L-Lyxose & No & $-2.232$ & $0.001$ \\ 
L-Xylulose & Yes & $-2.232$ & $0.001$ \\ 
CDP diacylglycerol & Yes & $-2.230$ & $0.001$ \\ 
\hline \\[-1.8ex] 
\end{tabular} 
\caption{
\footnotesize
Top 5 over-represented metabolites in the metabolic networks of taxa that receive the largest positive entries on variable 2. The column descriptions are provided with Table~S1.
}
\end{table} 

\begin{table}[!htbp]
\centering
\footnotesize
\begin{tabular}{@{\extracolsep{0pt}} lccc} 
\\[-1.8ex]\hline 
Metabolite & Synthesized & Enrich. Score & Adj. P\\ 
\hline \\[-1ex] 
Decaprenyl diphosphate & Yes & $2.300$ & $0.001$ \\ 
Acetyl-cystine-bimane & Yes & $2.191$ & $0.001$ \\ 
Bimane & No & $2.191$ & $0.001$ \\ 
Bimane conjugated mycothiol & Yes & $2.191$ & $0.001$ \\ 
Cys-1D-myo-inositol 2-deoxy-D-glucopyranoside & Yes & $2.191$ & $0.001$ \\ 
\hline \\[-1.8ex] 
\end{tabular} 
\caption{
\footnotesize
Top 5 over-represented metabolites in the metabolic networks of taxa that receive the largest positive entries on variable 2. The column descriptions are provided with Table~S1.
}
\end{table}

\begin{table}[!htbp] 
\centering 
\footnotesize
\begin{tabular}{@{\extracolsep{5pt}} lccc} 
\\[-1.8ex]\hline 
Metabolite & Synthesized & Enrich. Score & Adj. P\\ 
\hline \\[-1ex]
Corrinoid Iron sulfur protein & Yes & $3.093$ & $0.026$ \\ 
Methylcorrinoid iron sulfur protein & Yes & $3.093$ & $0.026$ \\ 
Citrate & No & $2.575$ & $0.038$ \\ 
L-Cystathionine & No & $2.468$ & $0.024$ \\ 
Indole & No & $2.459$ & $0.011$ \\ 
\hline \\[-1.8ex] 
\end{tabular}
\caption{
\footnotesize
Top 5 over-represented metabolites in the metabolic networks of taxa that receive the largest positive entries on variable 3. The column descriptions are provided with Table~S1.
}
\end{table} 

\begin{table}[!htbp] \centering 
\centering 
\footnotesize
\begin{tabular}{@{\extracolsep{5pt}} lccc} 
\\[-1.8ex]\hline 
Metabolite & Synthesized & Enrich. Score & Adj. P\\ 
\hline \\[-1ex]
Delta(1)-Piperideine-2-carboxylate & Yes & $-3.788$ & $0.0003$ \\ 
2 Oxoadipate C$_6$H$_6$O$_5$ & Yes & $-3.788$ & $0.0003$ \\ 
L 2 Aminoadipate C$_6$H$_10$NO$_4$ & Yes & $-3.788$ & $0.0003$ \\ 
L 2 Aminoadipate 6 semialdehyde & Yes & $-3.788$ & $0.0003$ \\ 
L-pipecolic acid & Yes & $-3.788$ & $0.0003$ \\
\hline \\[-1.8ex] 
\end{tabular}
\caption{
\footnotesize
Top 5 over-represented metabolites in the metabolic networks of taxa that receive the most negative entries on variable 4. The column descriptions are provided with Table~S1.
}
\end{table} 

\begin{table}[!htbp] \centering 
\centering 
\footnotesize
\begin{tabular}{@{\extracolsep{5pt}} lccc} 
\\[-1.8ex]\hline 
Metabolite & Synthesized & Enrich. Score & Adj. P\\ 
\hline \\[-1ex]
Riboflavin C$_{17}$H$_{20}$N$_4$O$_6$ & No & $-2.137$ & $0.001$ \\ 
L-Histidine & No & $-2.124$ & $0.001$ \\ 
L-Arginine & No & $-2.043$ & $0.001$ \\ 
L-Isoleucine & No & $-2.026$ & $0.001$ \\ 
L-Valine & No & $-2.017$ & $0.001$ \\ 
\hline \\[-1.8ex] 
\end{tabular}
\caption{
\footnotesize
Top 5 over-represented metabolites in the metabolic networks of taxa that receive the most negative entries on variable 8. The column descriptions are provided with Table~S1.
}
\end{table} 

\begin{table}[!htbp] \centering 
\centering 
\footnotesize
\begin{tabular}{@{\extracolsep{5pt}} lccc} 
\\[-1.8ex]\hline 
Metabolite & Synthesized & Enrich. Score & Adj. P\\ 
\hline \\[-1ex]
L-Histidine & No & $-1.791$ & $0.002$ \\ 
Tetradecenoyl-CoA & No & $-1.782$ & $0.002$ \\ 
Hexadecenoyl-CoA & No & $-1.766$ & $0.002$ \\ 
L-Arginine & No & $-1.765$ & $0.002$ \\ 
L-Threonine & No & $-1.702$ & $0.002$ \\ 
\hline \\[-1.8ex] 
\end{tabular}
\caption{
\footnotesize
Top 5 over-represented metabolites in the metabolic networks of taxa that receive the most negative entries on variable 10. The column descriptions are provided with Table~S1.
}
\end{table}

\end{document}